\documentclass[%
 reprint,
superscriptaddress,
 amsmath,amssymb,
 aps,
]{revtex4-2}

\usepackage{graphicx}% Include figure files
\usepackage{dcolumn}% Align table columns on decimal point
\usepackage{bm}% bold math
\usepackage{xcolor}
\begin{document}

\preprint{APS/123-QED}

\title{Precision Enhancement in Quantum Spatial Measurement by Squeezing-assisted Weak Value Amplification }
\author{Chaoxia Zhang}%
\affiliation{
State Key Laboratory of Quantum Optics and Quantum Optics Devices, Shanxi University, Taiyuan, 030006, China%\\This line break forced% with \\
}%
\affiliation{
College of Physics and Electronic Engineering, Shanxi University,Taiyuan, 030006, China%\\This line break forced with \textbackslash\textbackslash
}%
\author{Yongchao Chen}%
\affiliation{
State Key Laboratory of Quantum Optics and Quantum Optics Devices, Shanxi University, Taiyuan, 030006, China%\\This line break forced% with \\
}%
\affiliation{
College of Physics and Electronic Engineering, Shanxi University,Taiyuan, 030006, China%\\This line break forced with \textbackslash\textbackslash
}%
\author{Gang Chen}%
\affiliation{
State Key Laboratory of Quantum Optics and Quantum Optics Devices, Shanxi University, Taiyuan, 030006, China%\\This line break forced% with \\
}%
\affiliation{
Collaborative Innovation Center of Extreme Optics, Shanxi University, Taiyuan, 030006, China%\\This line break forced with \textbackslash\textbackslash
}%
\author{Hengxin Sun}%
\affiliation{
State Key Laboratory of Quantum Optics and Quantum Optics Devices, Shanxi University, Taiyuan, 030006, China%\\This line break forced% with \\
}%
\affiliation{
College of Physics and Electronic Engineering, Shanxi University,Taiyuan, 030006, China%\\This line break forced with \textbackslash\textbackslash
}%

\affiliation{
Collaborative Innovation Center of Extreme Optics, Shanxi University, Taiyuan, 030006, China%\\This line break forced with \textbackslash\textbackslash
}%
\author{Jing Zhang}%
\affiliation{
State Key Laboratory of Quantum Optics and Quantum Optics Devices, Shanxi University, Taiyuan, 030006, China%\\This line break forced% with \\
}%
\affiliation{
College of Physics and Electronic Engineering, Shanxi University,Taiyuan, 030006, China%\\This line break forced with \textbackslash\textbackslash
}%

\affiliation{
Collaborative Innovation Center of Extreme Optics, Shanxi University, Taiyuan, 030006, China%\\This line break forced with \textbackslash\textbackslash
}%
\author{Kui Liu}%
\email{liukui@sxu.edu.cn}
\affiliation{
State Key Laboratory of Quantum Optics and Quantum Optics Devices, Shanxi University, Taiyuan, 030006, China%\\This line break forced% with \\
}%
\affiliation{
Collaborative Innovation Center of Extreme Optics, Shanxi University, Taiyuan, 030006, China%\\This line break forced with \textbackslash\textbackslash
}%
\author{Rongguo Yang}%
\email{yrg@sxu.edu.cn}.
\affiliation{
State Key Laboratory of Quantum Optics and Quantum Optics Devices, Shanxi University, Taiyuan, 030006, China%\\This line break forced% with \\
}%
\affiliation{
College of Physics and Electronic Engineering, Shanxi University,Taiyuan, 030006, China%\\This line break forced with \textbackslash\textbackslash
}%

\affiliation{
Collaborative Innovation Center of Extreme Optics, Shanxi University, Taiyuan, 030006, China%\\This line break forced with \textbackslash\textbackslash
}%
\author{Jiangrui Gao}%
\affiliation{
State Key Laboratory of Quantum Optics and Quantum Optics Devices, Shanxi University, Taiyuan, 030006, China%\\This line break forced% with \\
}%

\affiliation{
Collaborative Innovation Center of Extreme Optics, Shanxi University, Taiyuan, 030006, China%\\This line break forced with \textbackslash\textbackslash
}%

\begin{abstract}
The precision enhancement is demonstrated in an optical spatial measurement based on weak value amplification (WVA) system and split-like detection, by injecting a TEM$_{10}$ squeezed vacuum beam. It is the first time to experimentally realize high-precision optical spatial measurement beyond the shot noise limit by using squeezing-assisted WVA. Based on the WVA technique, which can amplify the signal by increasing the number of photons that injected into the interferometer, squeezed beam injection can reduce the noise level and can further improve the signal-to-noise ratio(SNR). As a result, a SNR improvement of 2dB, i.e., 1.3 time precision enhancement, can be achieved,  and the obtained  displacement and tilt  sensitivity are $4.96{fm}/{\sqrt{H\text{z}}}\;$and $0.39{prad}/{\sqrt{H\text{z}}}\;$, respectively, by using a 2dB squeezed beam injection and 2.6\% postselection probability in WVA process. Our work provides an effective method to accomplish higher precision in quantum spatial measurement, which has potential applications in gravitational wave interferometer calibration, super-resolution quantum imaging, etc.
\end{abstract}

\maketitle

%\tableofcontents

\section{\label{sec:introduction}introduction}
Tilt and displacement are important attitude parameters because their measurement can be widely applied in many facets, such as engineering machinery alignment\cite{TONG2010Application}, spatial displacement tracking of vibrating structure\cite{huang2023spatial}, and land subsidence detection\cite{Yang1997MeasurementOG}, etc. Quantum precision measurement demonstrates the advantages of quantum mechanics, for example, in optical tilt and displacement measurements, which can be applied in biological measurement \cite{taylor2013biological}, cantilever displacement in atomic force microscopy\cite{meyer1988novel,pooser2015ultrasensitive}, high-resolution quantum imaging\cite{delaubert2008quantum,kolobov1999spatial,yang2016far}, high-quality astronomical observation \cite{kong2022fast}, weak
absorption measurements\cite{dickmann2019key} and ultra-precision calibration in gravitational wave detection \cite{morrison1994automatic,abbott2009ligo,scientific2017gw170104}.\\
\indent In experiment, increasing the average photon number is a conventional way to enhance the measurement precision, to approach the measurement limit. However, the intensity of the detected light might be limited by the detector saturation and the precision enhancement might be reduced or even eliminated. To solve this, weak value amplification (WVA) technology can be taken into account, which devotes to amplify small physical quantities through the postselection and can not only overcome the technical noise \cite{jordan2014technical} but also circumvent the problem of detector saturation\cite{xu2020approaching}. Abundant experiments demonstrated that WVA technique can improve precision by amplifying physical quantities, such as optical beam deflections\cite{dixon2009ultrasensitive,hosten2008observation,hogan2011precision,pfeifer2011weak,turner2011picoradian}, velocities\cite{viza2013weak}, phase shifts\cite{feizpour2011amplifying,jayaswal2014observing,xu2013phase,lupu2022negative}, frequency shifts\cite{starling2010precision}, and so on. Recently, our group experimentally accomplished a small-tilt measurement with a precision of 3.8nrad, based on the WVA technique\cite{zhang2023small}.  Although WVA technique performs remarkable in the above-mentioned facets, its measurement precision is still limited by the noise level.\\
\indent Squeezed light, which enables measurements beyond the shot noise limit, is a natural choice to combat noise in quantum precision measurement. Variable kinds of squeezed lights have been used in detecting magnetic field, phase, time, displacement, and gravitational wave, etc., to improving the signal-to-noise ratio(SNR). For small tilt and displacement measurement, the high-order-mode or spatial squeezed light can be used to improve precision due to their transverse spatial property. Precision enhancements of 1.5 times (from 2.3Å to 1.6Å) \cite{treps2003quantum} and 1.2 times (from 1.17Å to 0.99Å, by our group) \cite{sun2014experimental} are achieved in small-displacement measurements, by using spatial squeezed lights of 3.3dB and 2.2dB, respectively. Recently, the SNR of spatial tilt and displacement measurements were improved by 10 and 8.6dB, respectively, by using high-order spatially  squeezed light, by our group\cite{li2022higher}. It is noted that all these measurements were implemented in high frequency band, due to the inextricable technical noise in low frequency band. \\
\indent In this letter,  squeezed light and WVA technology are combined for the first time in the small-tilt measuring system, giving full play to their respective advantages to improve the measurement precision experimentally. It is found that the simultaneous use of WVA and squeezed beam can not only increase the number of photons while avoiding detector saturation but also reduce quantum noise, achieving complementary advantages. This work provides an effective method for ultra-precision measurement of small tilt and might have special applications in aerospace and gravitational wave detection.

\section{Experimental process and theoretical derivation}
\begin{figure}[h!]
\centering\includegraphics[width=8cm]{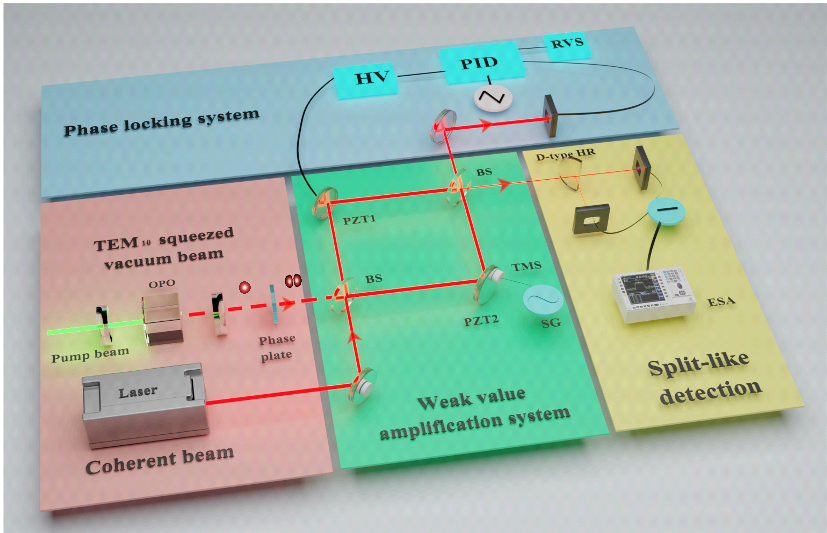}
\caption{Experimental setup. PZT1, PZT2: piezoelectric transducers, PID: proportional integral derivative controller,  HV: high voltage amplifier, RVS: reference voltage source,
SG: signal generator,  TMS: tilt modulation system,  D-type HR: D-type high reflection, ESA: electronic spectrum analyzer}
\end{figure}
 \indent The experimental setup is shown in Fig.1. A 1064nm coherent beam and a TEM$_{10}$ squeezed vacuum beam, as input part (shown in pink), enter the WVA system (shown in green). Two piezoelectric transducers are placed in two paths of the interferometer, respectively, namely PZT1 and PZT2. PZT1 is used to adjust the relative phase of the two paths. PZT2 is glued on the back top of the high-reflection mirror whose bottom is fixed, thus a one-dimensional tilt can be induced in the so-called tilt modulation system. The output beam from the bright port is detected by the phase locking system shown in blue, for controlling the relative phase of the two paths of the interferometer. At the dark port, half of the output beam transmits directly and the other half is reflected by the D-type high reflection mirror, then both enter the balanced homodyne detector, and the output signal can be analyzed by an electronic spectrum analyzer (ESA). Note that the attitude of the D-type high-reflection mirror must match the injecting squeezed TEM$_{10}$ mode, i.e., its edge must parallel to the phase-flip axis. We call this detection system shown by the yellow part in Fig.1, split-like detection, here our split-like detection system is composed of a D-type mirror, a pair of FD50 photodiodes (Fermionics, 50MHz bandwidth), a MODEL2117 detector (New Focus, 1-30000 adjustable gain). It is worth mentioned that this split-like detection system composed of a D-type high reflection mirror and a balanced homodyne detector has higher detection efficiency than the traditional SD. Therefore, it not only avoids the large loss caused by the inevitable element gap of conventional split detector(SD), but also has advantages of larger bandwidth and higher gain.  From this detection system, the beam tilt induced by PZT2 can be obtained, since there is a certain geometric relation between tilt $\theta$ and displacement $d$ ( $d=l\sin\theta $, where $l$ is the distance from the incident point of light beam to the fixed point of PZT2). \\
\indent According to the experimental setup shown in Fig. 1, matrix algebra can be used to describe the whole measurement process, especially the evolution of the two paths including transverse distribution and squeezed beam injection. At the beginning, the TEM$_{00}$ coherent beam and the TEM$_{10}$ squeezed vacuum beam are prepared and the electric fields of the two input ports of the Mach-Zehnder interferometer can be expressed as:
\begin{align}
  & {{{\mathbf{\hat{E}}}}_{\text{in }}}(x)=i{{(\frac{\hbar \omega }{2{{\varepsilon }_{0}}cT})}^{\frac{1}{2}}}\left
( \begin{array}{*{35}{l}}
   {{{\hat{a}}}_{0}}{{u}_{0}}(x)  \\
   {{{\hat{a}}}_{1}^{0}}u_{1}^{0}(x){{e}^{-i\psi}}  \\
\end{array} \right),  
 &  
\end{align} 
where $\hbar $ is the Planck constant, $\omega$ is the frequency of the light field, $c$ is the speed of light, $T$ is the detection integration time, ${{\varepsilon }_{0}}$ is the permittivity of the free space. ${{\hat{a}}_{0}}$ and  ${{\hat{a}}_{1}^{0}}$ represent the annihilation operators of the TEM$_{00}$ coherent state and the TEM$_{10}$ squeezed vacuum beam, respectively. ${{u}_{0}}(x)$ and $u_{1}^{0}(x)$ denote the transverse amplitude distributions of the TEM$_{00}$ coherent beam (pointer state) and the TEM$_{10}$ squeezed vacuum beam in one dimension, respectively. $\psi$ represents the relative phase between two input light fields of the interferometer. These electric fields enters the Mach-Zehnder interferometer through a 50:50 beamsplitter (BS), which can be expressed by a BS matrix 
\begin{equation}
\mathbf{B}=\frac{1}{\sqrt{2}}\left( \begin{array}{*{35}{l}}
   1 & i  \\
   i & 1  \\
\end{array} \right)
\end{equation}
\indent Then relative phase $\phi$ of the two paths of the interferometer and transverse momentum $k$ are introduced by PZT1 and PZT2, which can be described by a interaction matrix
\begin{equation}
\mathbf{M}=\left( \begin{matrix}
   {{e}^{i({kx}\;+{\phi }/{2}\;)}} & 0 \\
   0 & {{e}^{-i({kx}\;+\phi /2))}}  \\
\end{matrix} \right)
\end{equation}
\indent Therefore, the output electric fields of the  Mach-Zehnder  interfereometer can be expressed as 
\begin{equation}
\begin{split}
 &{{\mathbf{\hat{E}}}_{\text{out }}}(x)=\mathbf{BMB}{{{\mathbf{\hat{E}}}}_{\text{in }}}(x)\\
 &=A\left( \begin{matrix}
   \sin (kx+\phi /2){{{\hat{a}}}_{0}}{{u}_{0}}(x)+\cos (kx+\phi /2){{{\hat{a}}}_{1}^{0}}{{e}^{-i\psi}}u_{1}^{0}(x)  \\
   \cos (kx+\phi /2){{{\hat{a}}}_{0}}{{u}_{0}}(x)-\sin (kx+\phi /2){{{\hat{a}}}_{1}^{0}}{{e}^{-i\psi }}u_{1}^{0}(x)  \\
\end{matrix} \right)\\
\end{split}
\end{equation}
where $A=i{{(\frac{\hbar \omega }{2{{\varepsilon }_{0}}cT})}^{\frac{1}{2}}}$. Here the global phase of the output fields during the matrix multiplication is ignored.  Since the essence of weak value amplification is destructive interference, here only the dark port ($\phi$ is very small) of the Mach-Zehnder interferometer is focused on, meanwhile the output of the TEM$_{10}$ squeezed vacuum beam at this dark port is the constructive interference. For a very small $kx$, one can get $\sin (kx)\approx kx,\cos (kx)\approx 1$, by expanding the trigonometric functions to the first order. Then the electric field of the dark port can be expressed as
\begin{equation}
\begin{split}
 \mathbf{\hat{E}}_{_{\text{out }}}^{\text{dark}}(x)&=A\{\sin (\phi /2)\exp[{kx}\cot (\phi /2)]{{\hat{a}}_{0}}{{u}_{0}}(x)\\
&+\cos (\phi /2)\exp[{-kx}\tan (\phi /2)]{{\hat{a}}_{1}^{0}}{{e}^{-i\psi}}u_{1}^{0}(x)\}.
\end{split}
\end{equation}
 For Hermite-Gauss states, the following relations\cite{PhysRevApplied.13.034023,Delaubert2006TEM1H} are satisfied:
\begin{equation}
{{u}_{0}}(x)={{\omega }_{0}}{{u}_{1}}(x)/2x
\end{equation}
\begin{equation}
xu_{1}^{0}(x)=\frac{{{\omega }_{0}}}{2}\left( \sqrt{2}u_{2}^{0}(x)+u_{0}^{0}(x) \right)
\end{equation}
\begin{equation}
{{u}_{1}^{0}}{{(x)}_{k}}=exp(ikx){{u}_{1}^{0}}(x)={{u}_{1}^{0}}(x)+\frac{ik{{\omega }_{0}}}{2}\left(\sqrt{2}{{u}_{2}^{0}}(x)+{{u}_{0}^{0}}(x)\right).
\end{equation}
The annihilation operators ${{\hat{a}}_{0}}$ and ${{\hat{a}}_{1}^{0}}$ in Eq.(1) can be further expressed by ${{\hat{a}}_{0}}=\hat{X}+i\hat{Y}=\alpha +\delta {{\hat{a}}_{0}}=\sqrt{N}+\delta {{\hat{a}}_{0}}$ and $\hat{a}_{1}^{0}=\hat{X}_{1}^{0}{{e}^{-r}}+i\hat{Y}_{1}^{0}{{e}^{r}}=\delta \hat{X}_{1}^{0}{{e}^{-r}}+i\delta \hat{Y}_{1}^{0}{{e}^{r}}$(amplitude squeezing),  respectively. where $\hat{X},\hat{X}_{1}^{0}$ and $\hat{Y},\hat{Y}_{1}^{0}$ are the corresponding quadrature amplitude and phase operators, $\alpha$ and $\delta {{\hat{a}}_{0}}$ are the expectation value and the fluctuation, $N$ is the mean photon number, $r$ is the squeezing parameter. Considering all these relations, the output electric field of the dark port can be reformulated as
\begin{equation}
\begin{split}
 \mathbf{\hat{E}}_{_{\text{out }}}^{\text{dark}}(x)&\approx A[\sin (\phi /2)\sqrt{N}[{{u}_{0}}(x)+\frac{\cot {(\phi }/{2)}\;{{\omega }_{0}}k}{2}{{u}_{1}}(x)]\\
&+\cos (\phi /2)(\Delta\hat{X}_{1}^{0}{{e}^{-r}}+i\Delta\hat{Y}_{1}^{0}{{e}^{r}}){{e}^{-i\psi }}u_{1}^{0}(x)].
\end{split}
\end{equation}
\indent It is obvious that the tilt information is included in the induced $1$-th order mode (the second term), and the noise level depends on the original squeezed $1$-th order mode (the third term), i.e., the TEM$_{10}$ squeezed beam injection can reduce the noise level.  Note that $\cot (\phi /2)$ in the second term is equal to the weak value ${{A}_{w}}$ that defined by ${{A}_{w}}={\left\langle  f \right|\hat{A}\left| i \right\rangle }/{\left\langle  f \right|\left. i \right\rangle }\;={\cos (\phi /2)}/{\sin (\phi /2)=\cot (\phi /2)}\;$ in quantum description, which is equal to the classical description here\cite{howell2010interferometric}.  At the dark port,  $\phi$  is very very small, and $\cot (\phi /2)$  (the weak value amplification coefficient) is very very large. Therefore, the measured quantity tilt, which is related with $k$ ($k={2\pi \sin \theta }/{\lambda }\;$), can be amplified by $\cot (\phi /2)$ times by the weak value amplification system.   
The D-type mirror divides the output beam from the dark port into two identical parts in x axis (-$\infty \to $ 0 and $0 \to \infty$). Then, the photon numbers detected by the two detectors are ${{N}_{1}}=\frac{2{{\varepsilon }_{0}}cT}{\hbar \omega }\int_{-\infty }^{0}{\mathbf{\hat{E}}_{_{\text{out }}}^{dark}(x)\mathbf{\hat{E}}_{_{\text{out }}}^{\dagger dark}{{(x)}^{{}}}}dx$ and ${{N}_{2}}=\frac{2{{\varepsilon }_{0}}cT}{\hbar \omega }\int_{0}^{+\infty }{\mathbf{\hat{E}}_{_{\text{out }}}^{dark}(x)\mathbf{\hat{E}}_{_{\text{out }}}^{\dagger dark}(x)}dx$, respectively. According to the detection theory, the photon number difference ${\hat{N}}^{-}$ between two detectors is:\begin{equation}
\begin{split}
 & {{{\hat{N}}}^{-}}={{N}_{2}}-{{N}_{1}} \\ 
 & =A{{N}^{'}}\cot {(\phi }/{2)}\;{{\omega }_{0}}k+B\sqrt{{{N}^{'}}}(\delta \hat{a}_{1}^{0}{{e}^{-i\psi }}+\delta \hat{a}{{_{1}^{0}}^{\dagger }}{{e}^{i\psi }}) \\ 
 & =A{{N}^{'}}\cot {(\phi }/{2)}\;{{\omega }_{0}}k+B\sqrt{{{N}^{'}}}(\delta {{{\hat{X}}}_{1}}\cos \psi +\delta {{{\hat{Y}}}_{1}}\sin \psi ) \\
 & =\sqrt{{{N}^{'}}}(\sqrt{\frac{2}{\pi }}\sqrt{{{N}^{'}}}\cos \frac{\phi }{2}{{\omega }_{0}} k+\delta {{{\hat{X}}}_{1}}) \\ 
\end{split}
\end{equation}

where $A=\int_{0}^{+\infty }{{{u}_{1}}(x){{u}_{0}}(x)}dx-\int_{-\infty }^{0}{{{u}_{1}}(x){{u}_{0}}(x)}dx=\sqrt{{2}/{\pi }\;}$, $B=\int_{0}^{+\infty }{u_{_{1}}^{0}(x){{u}_{0}}(x)}dx-\int_{-\infty }^{0}{u_{_{1}}^{0}(x){{u}_{0}}(x)}dx=1$\cite{Delaubert2006TEM1H}. By locking the relative phase ($\psi =0$) of two input beams of the interferometer,  the quadrature amplitude squeezing can be achieved, i.e., only amplitude noise exists, $\left\langle {{\Delta }^{2}}{{{\hat{X}}}_{1}} \right\rangle ={{(\delta \hat{X}_{1}^{0}{{e}^{-r}})}^{2}}={{e}^{-2r}}$. 
The first term in the last raw of Eq.(10) is the signal part including the small-tilt information and the second term is the quantum noise. Consider that ${{N}^{'}}=N{{\sin }^{2}}({\phi }/{2}\;)$, in which $N$ is the number of photons injected into the bright port of the interferometer. Then, ${N}^{'}$ is the number of photons output from the dark port and can be detected by the split-like detection system.\\
\indent Then, the SNR of the tilt measurement is 
\begin{equation}
SNR=\frac{2}{\pi }{{\left( \frac{\sqrt{N}\cos \frac{\phi }{2}{{\omega }_{0}}k}{{{e}^{-r}}} \right)}^{2}}
\end{equation}
  As is known, the SNR is related with the measurement precision. The higher the SNR, the higher the measurement precision. When the relative phase $\phi$ is small enough, meanwhile the SNR becomes higher with the same waist, quantum fluctuation and photon number. In addition, it can also be seen from the above formula that the SNR  depend on the injected photon number $N$ of the interferometer, which means that effective beam tilt information can be obtained by postselecting only extremely small part of those tilted photons.\\
  
\section{Experimental results and analysis}
\indent In experiment,  we accomplish the optical spatial measurement based on WVA technique without and with squeezed light injection at low signal frequency (4kHz) and high signal frequency (500kHz).  As is known, a standard weak measurement process consists of three steps: preselection, weak interaction, and postselection. In our WVA system, the input and output ports of the Mach-Zehnder interferometer are corresponding to the preselection and postselection steps, respectively, while the weak interaction is from the tilt modulation system. At the dark port, destructive and constructive interference are happened to the TEM$_{00}$ mode and TEM$_{10}$ mode, respectively, therefore the weight of spatial information included in TEM$_{10}$ mode is amplified.\\
 \indent Firstly, the optimal postselection probability of WVA process is investigated, without squeezing assistance, and the noise power spectrum with different postselection probabilities for a 4kHz signal are shown in Fig.2. Here the output power of the weak value measurement system is fixed at $260\mu W$, and the injected optical power are chosen as $1mW,2mW,5mW$ and $10mW$ (the corresponding postselection probabilities are 26\%, 13\%, 5.2\%, 2.6\%), respectively. As is known, the postselection probability ${{p}_{f}}$ is the ratio of the output power ${{P}_{out}}$ to the input power ${{P}_{in}}$, i.e., ${{p}_{f}}={{{P}_{out}}}/{{{P}_{in}}}{=}{{\sin }^{2}}\frac{\phi }{2}$, where the relative phase $\phi$ is locked by PZT1 to keep constant output power in the dark port with different input power. From Fig.2, it is found that the smaller the postselection probability the larger the SNR, therefore,  $10mW$ input and the $260\mu W$ output power ($p_f=2.6\%$, shown by the purple curve) are chosen as the optimal experimental condition of the WVA process.\\

  \begin{figure}[h!]
\centering\includegraphics[width=7cm]{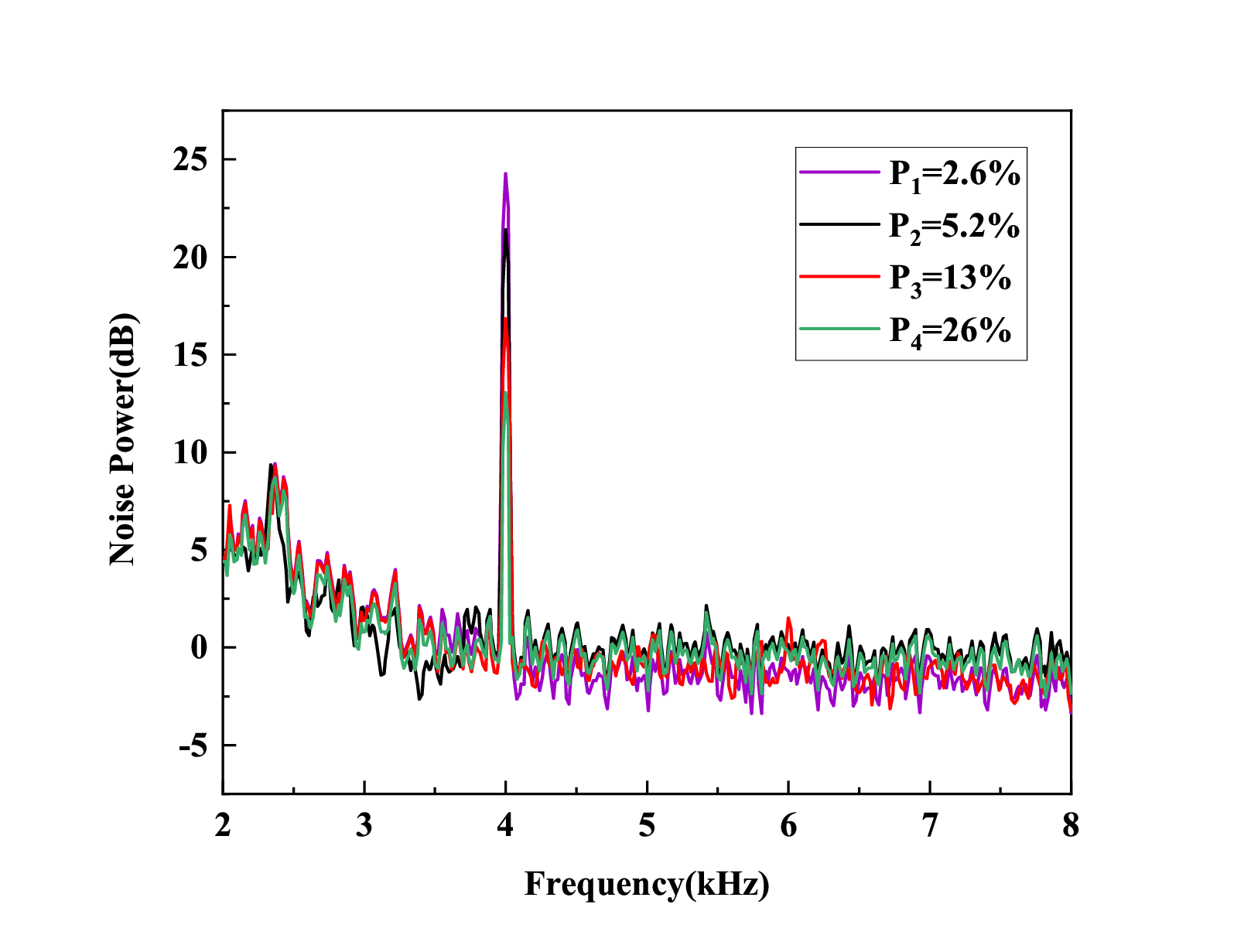}
\caption{Noise power spectrum with different postselection probabilities}
\end{figure}
\indent Then, to perform a quantum spatial measurement beyond the SNL, a TEM$_{10}$ squeezed vacuum beam is injected to the other input port of the interferometer, to fill the vacuum channel and reduce the noise. The TEM$_{10}$ squeezed vacuum beam is generated by letting a TEM$_{00}$ squeezed vacuum beam ( produced by OPO) go through a phase plate, as is shown in Fig.(1). So the TEM$_{10}$ squeezed vacuum beam here is actually an flipped mode\cite{Delaubert2006TEM1H}. At the dark port, the constructive interference happen for both the induced TEM$_{10}$ beam and the injected squeezed TEM$_{10}$ beam. 

\begin{figure}[htbp]
 \centering
 \begin{minipage}{0.48\linewidth}
  \centering
  \includegraphics[width=1.1\linewidth]{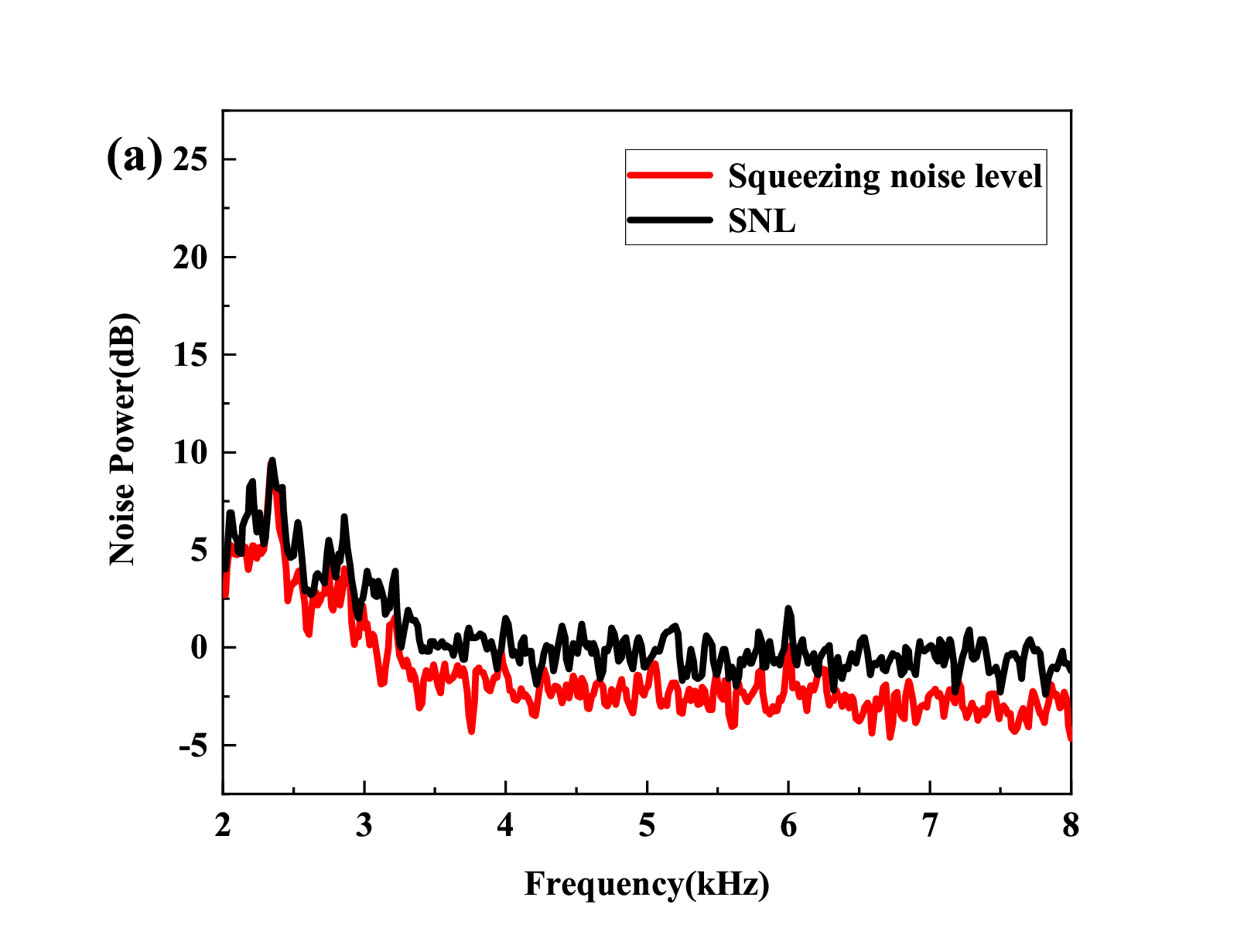}
  \label{chutian1}%文中引用该图片代号
 \end{minipage}
 \begin{minipage}{0.48\linewidth}
  \centering
  \includegraphics[width=1.1\linewidth]{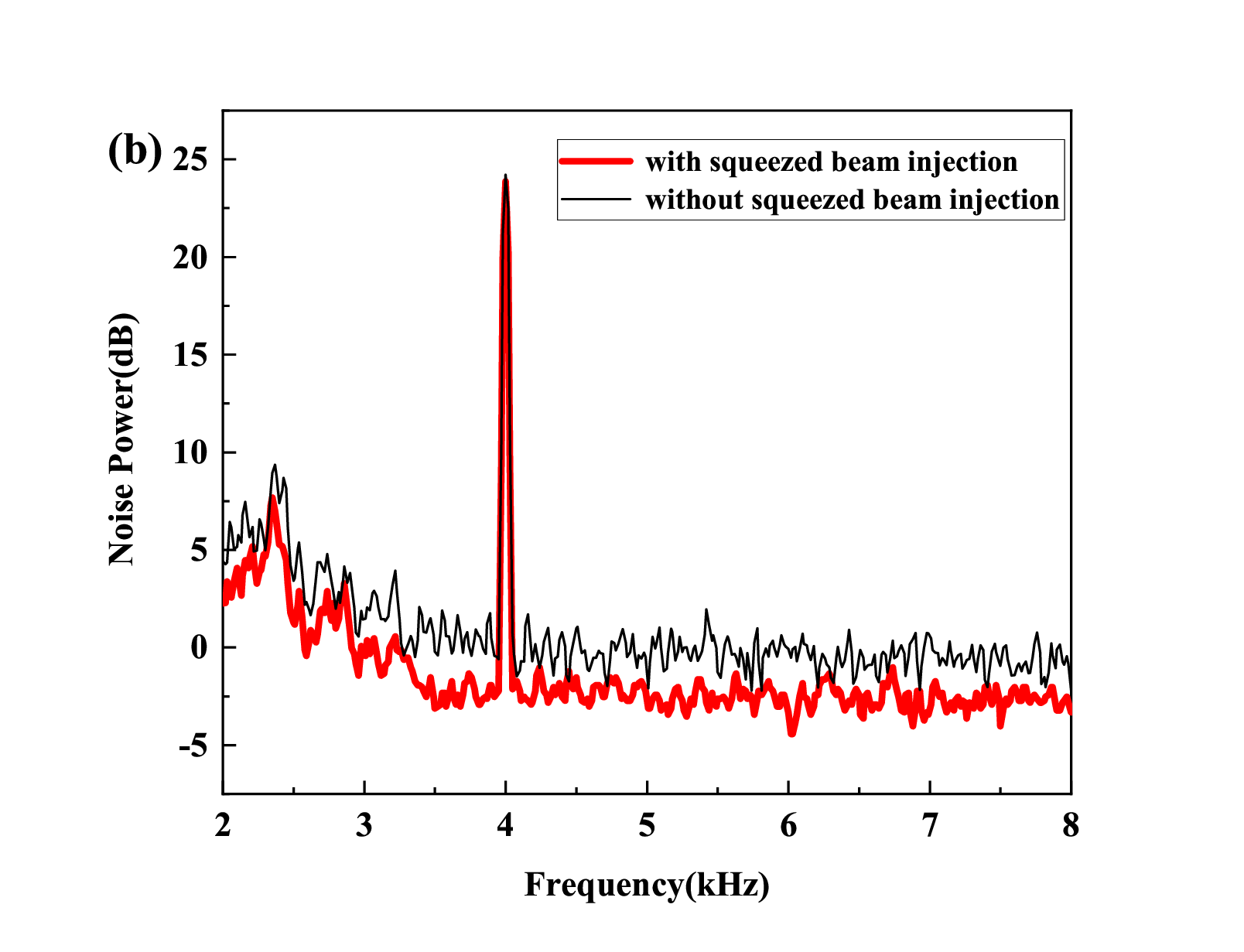}
  \label{chutian2}%文中引用该图片代号
 \end{minipage}.
 \centering
  \caption{(a) Squeezing level within the frequency range of 2-8 kHz. Black: the shot noise limit (SNL); Red: the squeezing noise level.   (b)Results of tilt measurement without and with squeezed beam injection.}
\end{figure}
\indent To characterize the squeezing level, the analysis frequency is scanned from 2kHz to 8kHz with the low-frequency spectrum analyzer(SR785,  STANFORD), as is shown in Fig.3(a). It is found that the generated squeezing is almost the same (2dB) in the frequency range from 3.5kHz to 8kHz for our OPO system. Here 4kHz is chosen as a signal frequency point for our spatial measurements at low frequency band. The experimental parameters are: the beam waist of TEM$_{00}$ ${{\omega }_{0}}\text{=1}\text{.86}mm$,  the resolution bandwidth RBW=1Hz. As is shown in Fig.3(b), based on our WVA system, SNR of 24dB (the black curve) and 26 dB (the red curve) are obtained without and with the squeezed light injection, respectively, i.e., the SNR can be improved 2dB at frequency 4kHz by using squeezed beam, for the same tilt signal. It is noted that this tilt signal is induced by the PZT2 drive voltage with the minimum output (1mV) of our signal generator. The measured small displacement at 4kHz is 0.10pm, and the corresponding tilt is 7.83prad. Corresponding to 3dB signal power, the minimum measurable displacement can reduced from 6.25fm ($6.25{fm}/{\sqrt{H\text{z}}}\;$) to 4.96fm ($4.96{fm}/{\sqrt{H\text{z}}}\;$) by injecting squeezed beam, and the corresponding minimum measurable tilt  is decreased from 0.49prad($0.49{prad}/{\sqrt{H\text{z}}}\;$) to 0.39prad ($0.39{prad}/{\sqrt{H\text{z}}}\;$). It is demonstrated that the enhancement of precision  is obtained in our optical spatial measurement based on WVA technique assisted with squeezed light.\\
\begin{figure}[htbp]
 \centering
 \begin{minipage}{0.48\linewidth}
  \centering
  \includegraphics[width=1.1\linewidth]{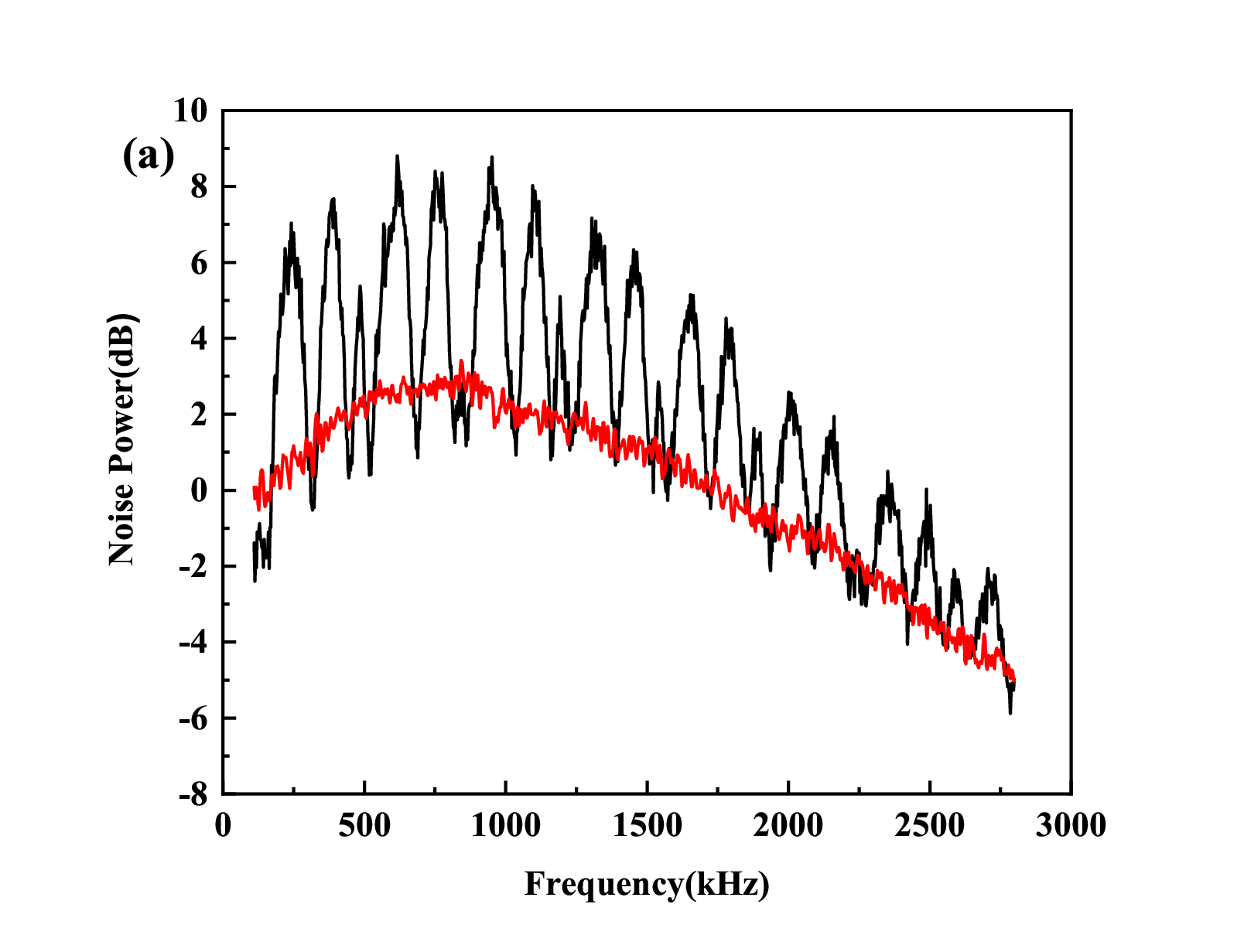}
  \label{chutian1}%文中引用该图片代号
 \end{minipage}
 \begin{minipage}{0.48\linewidth}
  \centering
  \includegraphics[width=1.1\linewidth]{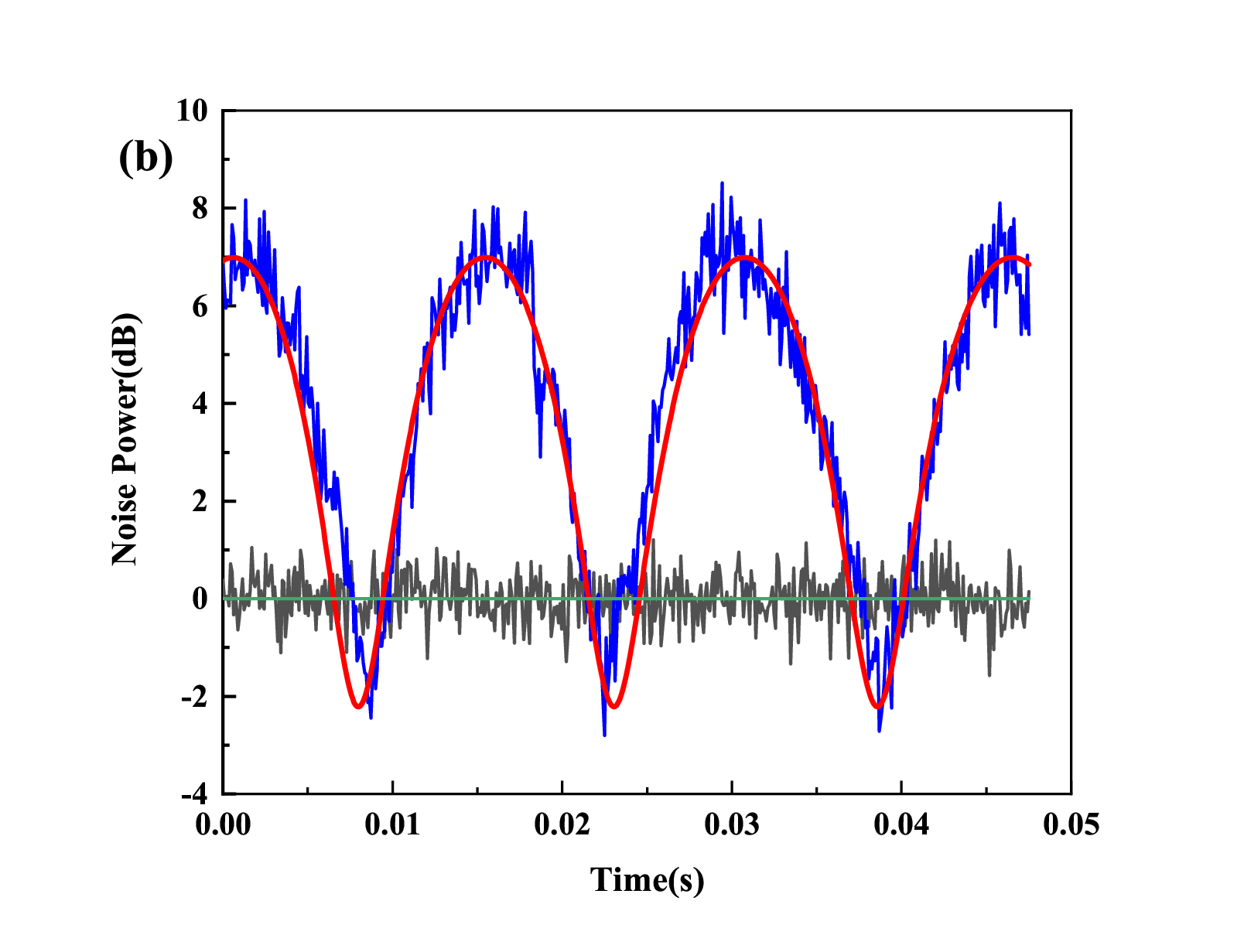}
  \label{chutian2}%文中引用该图片代号
 \end{minipage}.
 \centering
    \caption{(a) The observed squeezing noise levels versus the frequency, for detecting the squeezing of TEM$_{10}$ mode. Red : the shot noise limit (SNL); Black: the squeezing noise level.
    (b)The observed noise levels versus the sweeping time at 500kHz, for detecting the squeezing of TEM$_{10}$ mode. Black: the shot noise limit (SNL); blue: the local-phase-scanning noise level; red: the fit curve. }
\end{figure}

\indent In addition, the analysis frequency is scanned in a wide range of frequency by a high-frequency spectrum analyzer (Rohde \& Schwarz, FSVA3004), as is shown in Fig 4(a). It is found that the squeezing level is almost the same (2dB) in a wide frequency range (below 800kHz). The shot-noise level and the generated TEM$_{10}$ mode squeezing level with scanning the phase of the local beam at 500kHz are shown in Fig.4(b).\\
\begin{figure}[h!]
\centering\includegraphics[width=7cm]{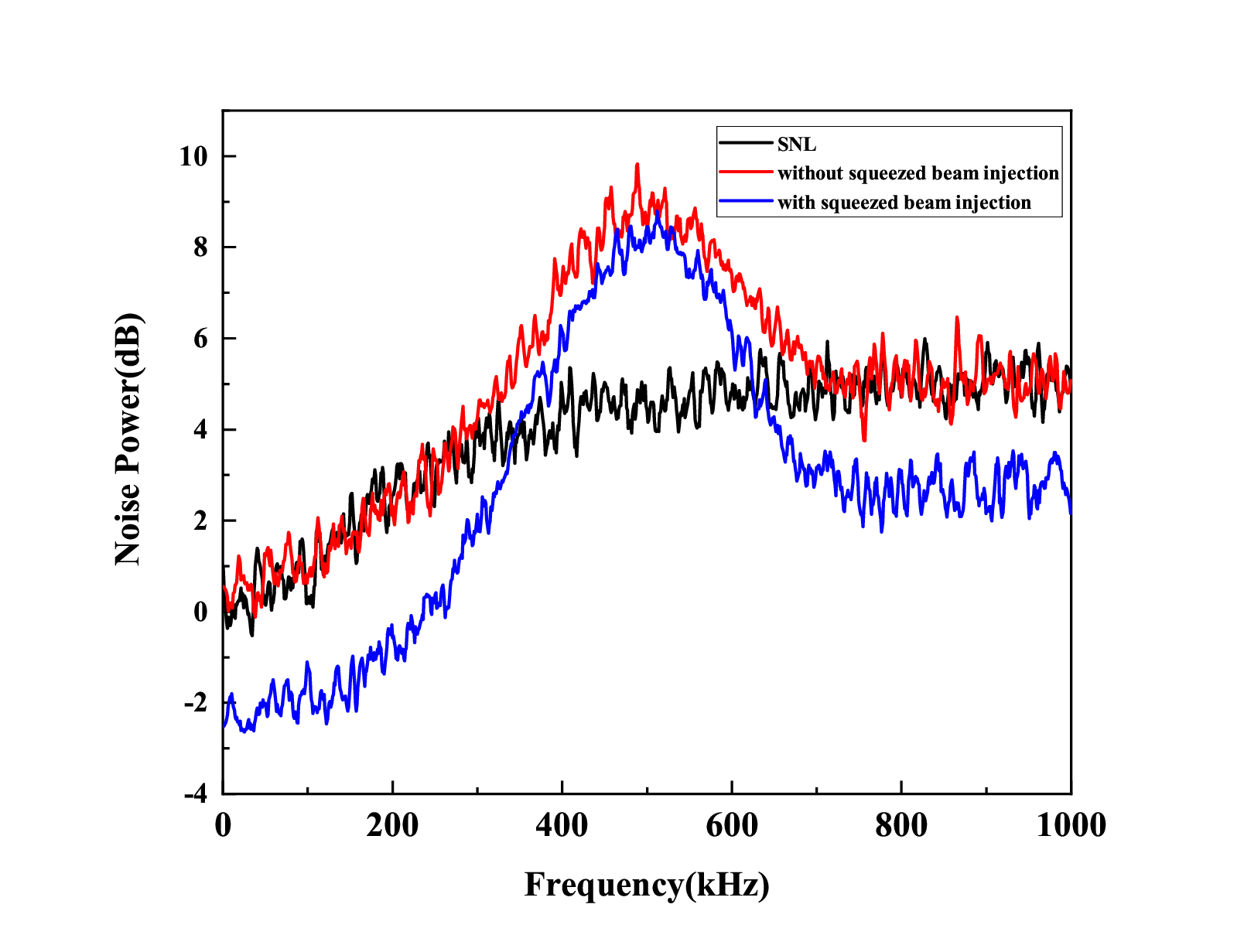}
\caption{Results of spatial measurements at a central frequency of 500 kHz. Black: the shot-noise-level without tilt modulation; red: tilt measurement performed with only coherent beam; blue: tilt measurement performed with both coherent and TEM$_{10}$ squeezed vacuum beam.}
\end{figure}
\indent 500kHz is chosen as a high frequency signal point for our spatial measurements. The experimental parameters are: beam waist of TEM$_{00}$ ${{\omega }_{0}}\text{=1}\text{.86}mm$, resolution bandwidth RBW=30kHz, input power $P{_{in}}$=10mW, output power ${{P}_{out}}$=260$\mu$W. The corresponding noise spectrum of the tilt measurement with and without the squeezed light injection are shown in Fig.5, respectively. A distinct peak centered at 500 kHz corresponding to the measured tilt signal can be found. The black curve shows the shot-noise-level. The red curve is achieved without squeezed beam injection. It is found that the signal peak at 500kHz is 3dB higher than the SNL, which means SNR=1. The minimum measurable displacement obtained from the measurement results is 1.08pm ($6.25{fm}/{\sqrt{H\text{z}}}\;$), and the corresponding minimum measurable tilt is 85prad($0.49{prad}/{\sqrt{H\text{z}}}\;$). The blue curve is obtained by adding the  TEM$_{10}$ squeezed vacuum beam and it is found that the SNR is improved by 2dB. And the corresponding minimum measurable displacement and tilt are 0.86pm( $4.96{fm}/{\sqrt{H\text{z}}}\;$), and 68prad( $0.39{prad}/{\sqrt{H\text{z}}}\;$), respectively.\\
\section{Conclusion}
\indent In summary, we experimentally complete the high-precision optical spatial measurement beyond the shot noise limit by using WVA technique (to amplify the signal and circumvent the detection saturation), squeezed light injection (to reduce the noise level) and split-like detection system (for large bandwidth, high gain, and small gap loss), and demonstrate the advantage of the WVA technique assisted with squeezed TEM$_{10}$ injection. The SNR is improved 2dB here, which can be further enhanced by optimizing the injecting squeezed light, i.e., generating the TEM$_{10}$ squeezed vacuum beam by OPO process directly, to avoid the loss of phase plate. Next, higher squeezing level at lower frequency band will under our consideration. This optical tilt and displacement measurement has potential applications in fiber optic gyro, optical waveguide sensor, super resolution imaging and inter-satellite positioning. In addition, this kind of measurement system also can be used to measure other small changes in various physical quantities.

\section*{Acknowledgments}
    This work was supported by National Key Research and Development Program of China (Grants No.2022YFA1404503, 2021YFC2201802); National Natural Science Foundation of China (Grants No.11874248, No.11874249, No.62027821 and  No.12074233).

\nocite{*}

\bibliography{apssamp}% Produces the bibliography via BibTeX.

\end{document}